\documentclass[preprint,aps,amsmath,nofootinbib,superscriptaddress]{revtex4}
\usepackage{graphicx}
\usepackage{bm}
\usepackage{epsfig}
\usepackage{pst-grad}

\newif\ifpdf
\ifx\pdfoutput\undefined
\pdffalse 
\else
\pdfoutput=1 
\pdftrue
\fi

\def\Dslash{D\!\!\!\!\slash}

\def\SppP{{\cal {P\!\!\!\!\hspace{0.04cm}\slash}}_\perp}

\def\nslash{n\!\!\!\slash}
\def\bnslash{\bar n\!\!\!\slash}

\def\Aslash{A\!\!\!\slash}
\def\OMIT#1{}

\newcommand{\nn}{\nonumber} 

\newcommand{\bn}{{\bar n}}
\newcommand{\bea}{\begin{eqnarray}}
\newcommand{\eea}{\end{eqnarray}}

\newcommand{\bnP}{\bar {\cal P}}
\newcommand{\ppP}{{\cal P}_\perp}
\newcommand{\bnPd}{\bar {\cal P}^{\raisebox{0.8mm}{\scriptsize$\dagger$}} }
\newcommand{\cP}{{\cal P}}

\newcommand{\mcdot}{\!\cdot\!}

\begin{document}
\ifpdf
\DeclareGraphicsExtensions{.pdf, .jpg}
\else
\DeclareGraphicsExtensions{.eps, .jpg}
\fi


\preprint{ \vbox{\hbox{INT-PUB-02-38} \hbox{UCSD/PTH 02-11}  
}}

\title{\phantom{x}\vspace{0.5cm} 
Power Counting in the Soft-Collinear Effective Theory
\vspace{0.5cm} }

\author{Christian W.~Bauer}
\affiliation{Department of Physics, University of California at San Diego,
	La Jolla, CA 92093\footnote{Electronic address: bauer@physics.ucsd.edu,
	pirjol@bose.ucsd.edu}}

\author{Dan Pirjol}
\affiliation{Department of Physics, University of California at San Diego,
	La Jolla, CA 92093\footnote{Electronic address: bauer@physics.ucsd.edu,
	pirjol@bose.ucsd.edu}}

\author{Iain W. Stewart\vspace{0.4cm}}
\affiliation{Institute for Nuclear Theory,  University of Washington, Seattle, 
	WA 98195 \footnote{Electronic address: iain@phys.washington.edu}
	\vspace{0.5cm}}


\begin{abstract}
\vspace{0.5cm}
\setlength\baselineskip{18pt}

We describe in some detail the derivation of a power counting formula for the
soft-collinear effective theory (SCET).  This formula constrains which operators
are required to correctly describe the infrared at any order in the
$\Lambda_{\rm QCD}/Q$ expansion ($\lambda$ expansion).  The result assigns a
unique $\lambda$-dimension to SCET graphs solely from vertices, is gauge
independent, and can be applied independent of the process.  For processes with
an OPE the $\lambda$-dimension has a correspondence with dynamical twist.

\end{abstract}

\maketitle


\newpage

\setlength\baselineskip{15pt}

\section{Introduction}

The physics of processes involving very energetic particles can be described in
terms of a soft-collinear effective theory
(SCET)~\cite{bfl,bfps,cbis,bpssoft,bfprs}. This effective theory has a
systematic expansion in a small parameter $\lambda$, which depends on the
typical offshellness of the partons participating in the hard process. Denoting
the large energy scale with $Q$ and the typical transverse momentum of
constituent quarks and gluons relative to the collinear axis with $p_\perp$, the
small parameter is $\lambda^2 = p_\perp^2/Q^2$. The effective theory is
constructed such that physical amplitudes are expressed as a sum of operators
corresponding to a power series in $\lambda$. Typically for exclusive processes
one has $p_\perp\sim \Lambda_{\rm QCD}$, so that $\lambda = \Lambda_{\rm
QCD}/Q$.  The purpose of this paper is to give a detailed derivation of the
power counting formula presented in Ref.~\cite{bps}, which depends only on the
vertices of the diagram.  This is the analog of having powers of $1/m_Q$
explicit in vertices in the Heavy Quark Effective Theory~\cite{bbook}. For a
given set of SCET fields our formula can be applied independent of the process
or the choice of gauge. It determines which operators are needed at any desired
order in the power counting to determine the cross section or decay rate at that
order.  As applications we show how this formula greatly simplifies the power
counting of graphs and how it is used to count powers of $Q$ in matrix
elements. We also discuss how the $\lambda$-dimension of operators relates to
twist for processes with an OPE such as DIS.

The degrees of freedom in the effective theory have well-defined momentum
scaling. Short distance fluctuations are integrated out and appear in Wilson
coefficients, while long distance fluctuations are described by effective theory
fields. We refer to Refs.~\cite{cbis,bpssoft,bfprs} for a detailed description
of the theory, and to
Refs.~\cite{bfl,bfps,cbis,bpssoft,bps,bfprs,mmps,xian,chay} for examples of how
the theory can be used to prove factorization theorems, sum Sudakov double
logarithms, and study power corrections.  Here only the features of SCET that
are relevant for explaining the power counting are reviewed.

The effective theory fields we will discuss are collected in
Table~\ref{table_pc}, together with their momentum and $\lambda$ scaling. The
momenta are given in light-cone momentum coordinates $(p^+, p^-, p^\perp)$,
defined as $p^-= \bn\cdot p$, $p^+=n\cdot p$, where $n^2=\bn^2=0$ and $\bn\cdot
n=2$. For simplicity we restrict ourselves to processes where only a single
collinear direction $n$ is relevant. The fields include collinear quarks and
gluons ($\xi_{n,p}$, $A_{n,q}^\mu$), massless soft quarks and gluons ($q_{s}$,
$A_{s}^\mu$), and massless ultrasoft (usoft) quarks and gluons ($q_{us}$,
$A_{us}^\mu$).  For processes with heavy quarks of mass $m_Q\sim Q$, the large
perturbative mass scale can be factored out as in Heavy Quark Effective
Theory. Then for power counting purposes the heavy quark fields $h_v$ are
identical to either the soft or usoft light quark fields. In
Table~\ref{table_pc} the fields are assigned a scaling with $\lambda$ such that
the action for their kinetic terms is order $\lambda^0$, and we also list the
scaling of label operators and Wilson lines. Label operators $\bnP$ and
$\cP^\mu$ are introduced to facilitate the power counting and separation of
momentum scales~\cite{cbis}. Integrating out offshell fluctuations with $p^2\gg
(Q\lambda)^2$ induces the collinear Wilson line $W_n[\bn\mcdot A_{n,q}]$ and
soft Wilson line $S_n[n\mcdot A_s]$, which appear in a way that preserves gauge
invariance~\cite{bfps,bpssoft}. The usoft Wilson line $Y_n[n\mcdot A_{us}]$ is
induced by the field redefinitions $\xi_{n,p}=Y_n \xi^{(0)}_{n,p}$ and
$A_{n,q}=Y_n A_{n,q}^{(0)} Y_n^\dagger$ which decouple usoft gluons from the
leading order collinear Lagrangian~\cite{bpssoft}.
\begin{table}[t!]
\begin{center}
\begin{tabular}{clc|cc|cc}
\hline\hline
 & Type & Momenta $p^\mu=(+,-,\perp)$\hspace{0.4cm} 
   & Fields ($f$) & Scaling ($e^f$) \hspace{0.2cm} 
   & \hspace{0.2cm}Operator\hspace{0.2cm} & \hspace{0.2cm} Scaling \\ 
   \hline
 & collinear & $p^\mu\sim (\lambda^2,1,\lambda)$ 
   & $\xi_{n,p}$ & $\lambda$ & $\bnP$, $W_n$ & $\lambda^0$ \\
 &&& ($A_{n,p}^+$, $A_{n,p}^-$, $A_{n,p}^\perp$)\hspace{0.4cm} & 
  ($\lambda^2$,$1$,$\lambda$) & $\cP_\perp^\mu$ & $\lambda$ \\ \hline
 & soft &  $p^\mu\sim (\lambda,\lambda,\lambda)$ 
   & $q_{s,p}$ & $\lambda^{3/2}$ & $S_n$ & $\lambda^0$ \\
 & & & $A_{s,p}^\mu$ & $\lambda$ & $\cP^\mu$ & $\lambda$ \\ \hline 
 & usoft &  $k^\mu\sim (\lambda^2,\lambda^2,\lambda^2)$
   & $q_{us}$ & $\lambda^3$ & $Y_n$ & $\lambda^0$ \\
 & & & $A_{us}^\mu$ & $\lambda^2$  \\
\hline\hline
\end{tabular}
\end{center} \label{table_pc}
\caption{\setlength\baselineskip{12pt} Power counting for SCET momenta and
fields as well as momentum label operators and Wilson lines. 
\setlength\baselineskip{18pt}}
\end{table}

Interactions in SCET appear either in the effective theory action or in external
operators (which are often operators or currents generated by electromagnetic or
weak interactions).  It is useful to divide the full action of SCET into four
pieces
\begin{eqnarray}
 S = S^{U} + S^{S} + S^{C} + S^{SC}\,,
\end{eqnarray}
where $S^U$ has purely usoft interactions, $S^S$ contains interactions with one
or more soft fields, $S^C$ contains interactions with one or more collinear
fields, and $S^{SC}$ contain possible mixed soft-collinear terms. Once offshell
fluctuations are integrated out the mixed soft-collinear interactions typically
occur in external operators, however mixed collinear-usoft interactions do
appear in $S^C$.  The leading terms in $S^U$, $S^S$, and $S^C$ are the order
$\lambda^0$ kinetic terms for the usoft, soft, and collinear fields.

The leading order Lagrangians for (u)soft light quarks and gluons are the same
as QCD. For heavy quarks we have the HQET Lagrangian, ${\cal L}=\bar h_v\,
iv\cdot D\, h_v+\ldots$.  The collinear quark Lagrangian has an expansion in
$\lambda$,
\begin{eqnarray}
 {\cal L}_c &=&  {\cal L}_c^{(0)} + {\cal L}_c^{(1)} + {\cal L}_c^{(2)} 
   + \ldots \,.
\end{eqnarray}
The first three terms are 
\begin{eqnarray} \label{Lc}
{\cal L}_{c}^{(0)} 
 &=&   \bar\xi_{n,p'}\:  \bigg\{  i\, n\mcdot  D\!+\! g n\mcdot A_{n,q} 
  + \Big( \SppP\! + g \Aslash_{n,q}^\perp\Big)\, W \frac{1}{\bnP} W^\dagger\,
   \Big( \SppP\!  + g \Aslash_{n,q'}^\perp\Big) \bigg\}
  \frac{\bnslash}{2}\, \xi_{n,p}  \,, \\
{\cal L}_{c}^{(1)}
 &=&  \bar\xi_{n,p'}\bigg\{ i\Dslash_\perp W \frac{1}{\bnP} W^\dagger\,
   \big(\SppP\!+g\Aslash_{n,q'}^\perp\big) 
  + \Big( \SppP\! + g \Aslash_{n,q}^\perp\Big)\, W \frac{1}{\bnP} W^\dagger\,
   i\Dslash_\perp \bigg\} \frac{\bnslash}{2}\, \xi_{n,p} \,, \nn \\
{\cal L}_c^{(2)} 
 &=& \bar \xi_n \left\{ i\Dslash_\perp W\frac{1}{\bnP} W^\dagger i\Dslash_\perp 
 -\Big( \SppP\! + g \Aslash_{n,q}^\perp\Big)
W \frac{1}{\bnP} W^\dagger
(\bn\mcdot iD) W \frac{1}{\bnP} W^\dagger 
\Big( \SppP\! + g \Aslash_{n,q}^\perp\Big)
\right\}
\frac{\bnslash}{2}\xi_n \,, \nn
\end{eqnarray}
where $iD^\mu = i\partial^\mu + g A^\mu_{us}$. Here ${\cal L}_c^{(0)}$ gives the
order $\lambda^0$ interactions~\cite{bfps,cbis}. The expression for ${\cal
L}_c^{(1)}$ was first given in Ref.~\cite{chay}, and that for ${\cal L}_c^{(2)}$
in Ref.~\cite{mmps}.  

For any operator in SCET, a scaling $\lambda^k$ can be immediately assigned by
adding up the factors of $\lambda$ associated with the scaling of its fields,
derivatives, and label operators. In time-ordered products or Feynman diagrams
these operators appear as vertices. To power count these it is useful to
introduce indexes $V_k^i$ which count operators in a graph. Here $V_k^i$ counts
the number of operators that scale as $\lambda^k$ and are of type $i$, where the
type depends on the field content. The four vertex indexes we require are:
\begin{enumerate}
 \item[] $V_k^C$ for vertices involving only collinear and usoft
  fields ,
 \item[] $V_k^S$ for vertices involving only soft and usoft fields ,
 \item[] $V_k^{SC}$ for vertices with both soft and collinear fields ,
 \item[] $V_k^{U}$ for vertices with {\em only} usoft fields.
\end{enumerate}
Note the important point that the mixed soft-collinear operators require a
separate index. For Non-Relativistic QCD indexes analogous to these were defined
in Ref.~\cite{LMR}.  As an example of how these indexes work consider the purely
collinear DIS operator~\cite{bfprs}
\begin{eqnarray} \label{DIS}
 {\cal O}_1 = \frac{1}{Q}\: \bar\xi_{n,p'} W_n\: \frac{\bnslash}{2}\: 
  C(\bnP_+,\bnP_-,Q,\mu)\: W_n^\dagger\, \xi_{n,p}\,.
\end{eqnarray}
where $\bnP_\pm=\bnP^\dagger \pm \bnP$ and $C$ a Wilson coefficient.  When
taking the proton matrix element of ${\cal O}_1$, the $\bnP_+$ dependence of the
Wilson coefficient leads to a convolution involving the quark or antiquark
parton distribution functions.  Since $\xi\sim\lambda$ and $W_n\sim\lambda^0$
this operator scales as $\lambda^2$. Thus $k=2$, and since the operator only
involves collinear fields a single insertion of ${\cal O}_1$ makes the index
$V_2^C=1$.

Now consider an arbitrary loop graph built out of insertions of external
operators along with propagators and interactions from $S$. In the effective
theory each such graph scales with a unique power of $\lambda$, say
$\lambda^\delta$ for some $\delta$. Since the effective theory is constructed to
include all the relevant infrared degrees of freedom, the set of all SCET graphs
scaling as $\lambda^\delta$ will reproduce the infrared structure of QCD at this
order. Our goal is to prove that
\begin{eqnarray}\label{pc}
  \delta = 4u + 4 + \sum_k (k-4) \big( V_k^C + V_k^S + V_k^{SC}) + (k-8)
  V_k^{U}\,,
\end{eqnarray}
where $u=1$ if the graph is purely usoft and $u=0$ otherwise. We will refer to
Eq.~(\ref{pc}) as the {\em vertex power counting formula}. This result applies
to any physical process whose infrared structure can be described by the fields
in Table~\ref{table_pc}. In Ref.~\cite{bps}, this formula was given\footnote{The
formula in Ref.~\cite{bps} has a different overall offset than Eq.~(\ref{pc}),
since for the proof there only relative scalings mattered. In general there is
additional information in the normalization which is explored in this paper.},
but the steps in its derivation were not described.  Eq.~(\ref{pc}) expresses
the important result that the power of $\lambda$ associated with an arbitrary
diagram can be determined entirely by the scaling of operators at its vertices.
This is the analog of having all powers of $1/m_Q$ explicit in the vertices in
the Heavy Quark Effective Theory.

Only external operators can have $k<4$, such as in the DIS operator in
Eq.~(\ref{DIS}). Furthermore, physical considerations always limit the number of
external operator insertions (usually to just 1). For example, in DIS multiple
insertions of ${\cal O}_1$ would require multiple electromagnetic
interactions. Thus, at leading order in $\lambda$ only graphs built out of a
fixed number of external operators plus $V_4^C$, $V_4^S$, $V_4^{SC}$, and
$V_8^{U}$ vertices need to be included. These vertices are exactly those
described by the order $\lambda^0$ actions derived in
Refs.~\cite{bfps,cbis,bpssoft}.  At one higher order in $\lambda$ we only need
to add a {\em single} vertex with a higher power of $k$ to the vertices included
above. For instance, a single $V_5^C$.  From this discussion the utility of
Eq.~(\ref{pc}) for describing higher and higher orders in $\lambda$ should be
fairly evident.

\vspace{0.2cm}
\noindent{\bf Direct Power Counting and the Derivation of Eq.(5)}
\vspace{0.2cm}
 
The derivation of power counting formulae such as the one in Eq.~(\ref{pc}) has
a history back to Weinberg~\cite{wein} and Sterman~\cite{Sterman,Ellis}.
Our result in Eq.~(\ref{pc}) provides a simple way of determining the order in
$\Lambda_{\rm QCD}/Q$ of any given diagram. The part of our proof from
Eqs.~(\ref{raw}-\ref{pc2}) follows the concise approach of Ref.~\cite{LMR}
(where a power counting formula for Non-Relativistic QCD was derived).

An intuitive method of power counting diagrams involves counting powers of
$\lambda$ for the loop measures, propagators, vertices, and external lines. We
will refer to this as the ``direct'' method of power counting and use it as our
starting point. In the context of collinear interactions the direct method has
been employed with the threshold expansion in Ref.~\cite{BBNS2}, although
without counting powers of $\lambda$ for external fields. For this method it it
is more intuitive to begin with indexes $\tilde V_{k'}^j$ which are analogous to
the $V_k^j$, but do not include the scaling for the fields. Thus, $\tilde
V_{k'}^j$ directly count the scaling of the vertex Feynman rules.  For example,
an operator $\bar\xi_n\, \cP_\perp^2\, \xi_n$ would be $\tilde V_2^C=1$,
whereas $\bar\xi_n\, A_{n,q}^{\perp\,2} \,\xi_n$ is $\tilde V_0^C=1$.  Below we
will show that this direct power counting method can be reduced to
Eq.~(\ref{pc}). Readers not interested in the derivation can safely skip to
Application 1.

Consider an arbitrary graph containing $L^i$ loops and $I^i$ propagators of type
$i$ with $i\in \{C,S,U\}$, $\tilde V_k^j$ vertices of type $j$, and $E^f$
external lines of type $f$, where $f$ runs over the fields in
Table~\ref{table_pc}.  Counting the powers of $\lambda$ associated with the
vertices, loop measures, propagators, and external lines we find that the graph
scales as $\lambda^\delta$ with
\begin{eqnarray}\label{direct}
\delta = \sum_{k'} \: {k'} \big( \tilde V_{k'}^C \!+\! \tilde V_{k'}^S 
  \!+\! \tilde V_{k'}^{SC} \!+\! \tilde V_{k'}^{U}) + 4 L^C\! + 4 L^S\! 
  \! + 8 L^{U} - \eta_\alpha I^C\! -2 I^S\! -4 I^U 
  + \sum_f e^f E^f\,. \nn\\[-12pt]
\end{eqnarray}
which we refer to as the {\em direct power counting formula}.  To derive
Eq.~(\ref{direct}), consider an arbitrary time-ordered product of operators
$\sim \int\!  d^4x_i \exp(\cdots) \langle {\cal O}_1(x_1)\cdots {\cal
O}_n(x_n)\rangle$ and transform all loops and internal lines to momentum
space. The scaling of momenta in the vertices give the $\tilde V^j$ terms in
Eq.~(\ref{direct}).  With the momenta and fields scaling as in
Table~\ref{table_pc}, the integrations over loop momenta give the $L^i$ terms,
and the external fields give the $E^f$ terms, where $e^f$ is the appropriate
power of $\lambda$ from the table. The $E_f$ term is important if we wish to
compare the size of operators with different external lines. Finally, the
internal propagators give the $I^i$ terms. Since the scaling of collinear gluons
is not homogeneous, $(A_n^+,A_n^-,A_n^\perp)\sim (\lambda^2,\lambda^0,\lambda)$,
their propagator contribution is gauge and component dependent. This dependence
is encoded in the coefficient $\eta_\alpha$. In Feynman gauge $\eta_\alpha=2$
for all collinear particles, however in a general covariant gauge
$\eta_\alpha=2$ for fermions, but the gluon propagator
\begin{eqnarray} \label{AA}
   A^\mu_n\ A^\nu_n\ \ \Longrightarrow\ \ 
    \frac{-i}{p^2} \Big( g^{\mu\nu}+ \alpha \frac{p^\mu p^\nu}{p^2} \Big) \,,
\end{eqnarray}
gives $\eta_\alpha=\{0,1,2,2,3,4\}$ for the
$\{++,+\!\!\perp,+-,\perp\perp,-\!\!\perp, --\}$ components. We will see that
the $\eta_\alpha$ gauge dependence in Eq.~(\ref{direct}) is cancelled by a
similar dependence in $\tilde V^C_{k'}$.

To proceed we switch to the $V_k^i$ indexes. The difference between the two
types of vertex indexes is the powers of $\lambda$ associated with the fields in
an operator. For example, both $\bar\xi_n\, \cP_\perp^2\, \xi_n$ and
$\bar\xi_n\, A_{n,q}^{\perp\,2} \,\xi_n$ count as $V_4^C=1$.  Since the fields
in a vertex are either contracted with another field or correspond to an
external line we have
\begin{eqnarray} \label{switchV}
  \sum_{k,i} \: {k}\: V_{k}^i  = \sum_{k',i} \: {k'}\: \tilde V_{k'}^i 
   + (4 \!-\! \eta_\alpha) I^C + 2 I^S + 4 I^U +\sum_f e^f E^f \,.
\end{eqnarray}
For each internal line two fields are contracted and eliminated for a momentum
space propagator. The difference in scaling of the two fields and the propagator
induces the $I^i$ correction terms in Eq.~(\ref{switchV}). For instance two
collinear fermion fields scale as $\lambda^2$ on the LHS. If they are contracted
then a $(4-\eta_c)I^C=(4-2)I^C=2I^C$ term accounts for them on the RHS. The
$E^f$ terms account for them if they are not contracted. In a similar way, each
soft propagator or external line is also accounted for. The most non-trivial
contraction is that of two collinear gluon fields $A^\mu_n\ A^\nu_n$ since their
scaling is inhomogeneous. However, it is straightforward to see, for
example in a general covariant gauge, that these contractions are correctly
encoded by the $(4-\eta_\alpha)I^C$ term in Eq.~(\ref{switchV}).  Using
Eq.~(\ref{switchV}) to eliminate the $\tilde V_{k'}^i$'s from
Eq.~(\ref{direct}) leaves
\begin{eqnarray}\label{raw}
  \delta = \sum_k \: k \big( V_k^C + V_k^S + V_k^{SC}+ V_k^{U})
  + 4 L^C + 4 L^S + 8 L^{U} - 4 I^C -4 I^S -8 I^U \,,
\end{eqnarray}
whose terms are now explicitly gauge independent. This is made possible by the
fact that the $V_k^i$ indexes assign a homogeneous $\lambda$-dimension to gauge
invariant operators.

Eq.~(\ref{raw}) can be further simplified by using Euler topological identities,
which connect the number of loops, lines and vertices in an arbitrary
graph. For the complete graph we have
\begin{eqnarray} \label{E1}
 \mbox{(E1):\hspace{0.8cm}  }
 \sum_k (V_k^C + V_k^S + V_k^{SC}+ V_k^{U}) + (L^S + L^C + L^U) - 
 (I^S + I^C + I^U) = 1\,.\hspace{0.8cm}
\end{eqnarray}
Using this result to remove the $L^U-I^U$ factor from Eq.~(\ref{raw}) leaves
\begin{eqnarray} \label{raw2}
 \delta = 8+\sum_k \: (k-8) \big( V_k^C + V_k^S + V_k^{SC}+ V_k^{U})
  - 4 L^C - 4 L^S  + 4 I^C +4 I^S  \,. 
\end{eqnarray}
To proceed further we consider two cases, i) graphs with purely usoft fields,
and ii) graphs with $\ge 1$ soft or collinear field. In case i) $L^C=L^S=I^C=I^S
= V_k^C=V_k^S=V_k^{SC}=0$ and Eq.~(\ref{raw2}) gives the final result which is
$\delta=8 +\sum_k (k-8) V_k^U$.

In case ii) we can use the hierarchy in $p^2$ of the modes in Table I to derive
additional Euler relations by removing modes one at a time, starting with those
propagating over the longest distance~\cite{LMR}. This is possible since the
graph must stay connected when probed at the shorter distance scales.  Thus, by
removing all the usoft lines in the graph one can draw a reduced graph
containing only soft and collinear modes, also of course keeping vertices that
are not purely usoft.  The Euler identity for this reduced graph reads
\begin{eqnarray}
 \mbox{(E2):\hspace{2cm}  }
 \sum_k (V_k^C + V_k^S + V_k^{SC}) + (L^S + L^C ) - (I^S + I^C) = 1\,.
 \hspace{3cm}
\end{eqnarray}
It is easy to see that (E2) can be used to eliminate the remaining terms in
Eq.~(\ref{raw2}) that depend on the loop measures and internal lines. Because
the soft and collinear terms have the same prefactor, a single Euler relation
eliminates all of them simultaneously. This differs from NRQCD where removing
potential and soft factors require distinct topological
identities~\cite{LMR}. Using (E2) in Eq.~(\ref{raw2}) leaves the final result
[for graphs with at least one soft and/or collinear field, ie. case ii)],
\begin{eqnarray}\label{pc2}
  \delta = 4 + \sum_k (k-4) \big( V_k^C + V_k^S + V_k^{SC}) + (k-8) V_k^{U}\,.
\end{eqnarray}
Together cases i) and ii) reproduce Eq.~(\ref{pc}) which is the main result of
this paper.  It should be noted that the derivation of Eq.~(\ref{pc}) would not
be possible if offshell degrees of freedom had been retained since the power
counting for fields generating offshell fluctuations is
ambiguous~\cite{bpssoft}. In the remainder of the paper we illustrate how
Eq.~(\ref{pc}) can be used on a few examples of physical interest.

\vspace{0.2cm}
\noindent{\bf Application 1: A simplified power counting of graphs}
\vspace{0.2cm}

\begin{figure}[t!]
\centerline{ 
\includegraphics[width=1.in]{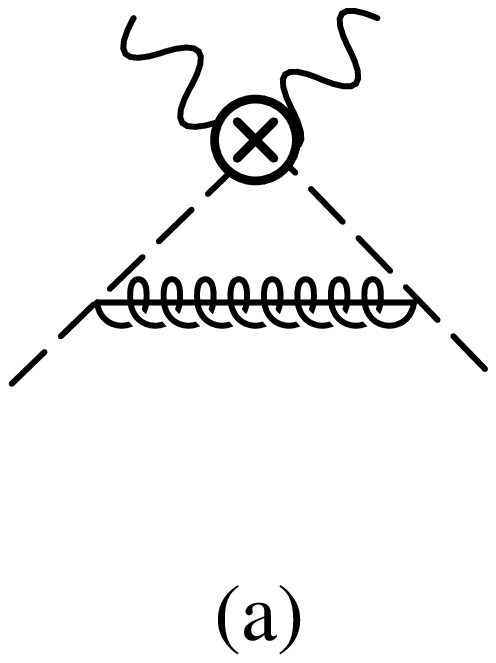}\hspace{1.3cm}
\includegraphics[width=2.in]{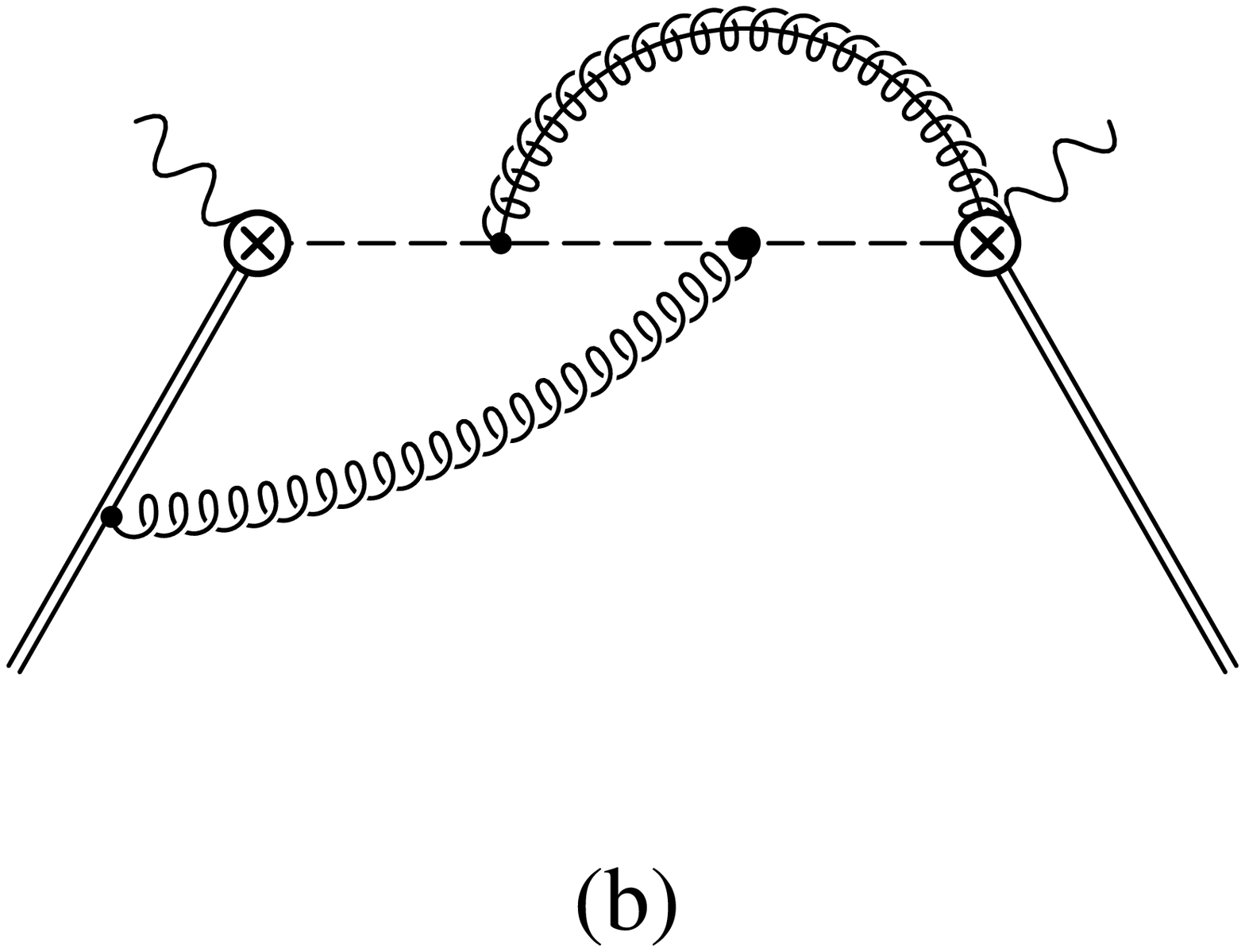}\hspace{1.3cm}
\includegraphics[width=1.5in]{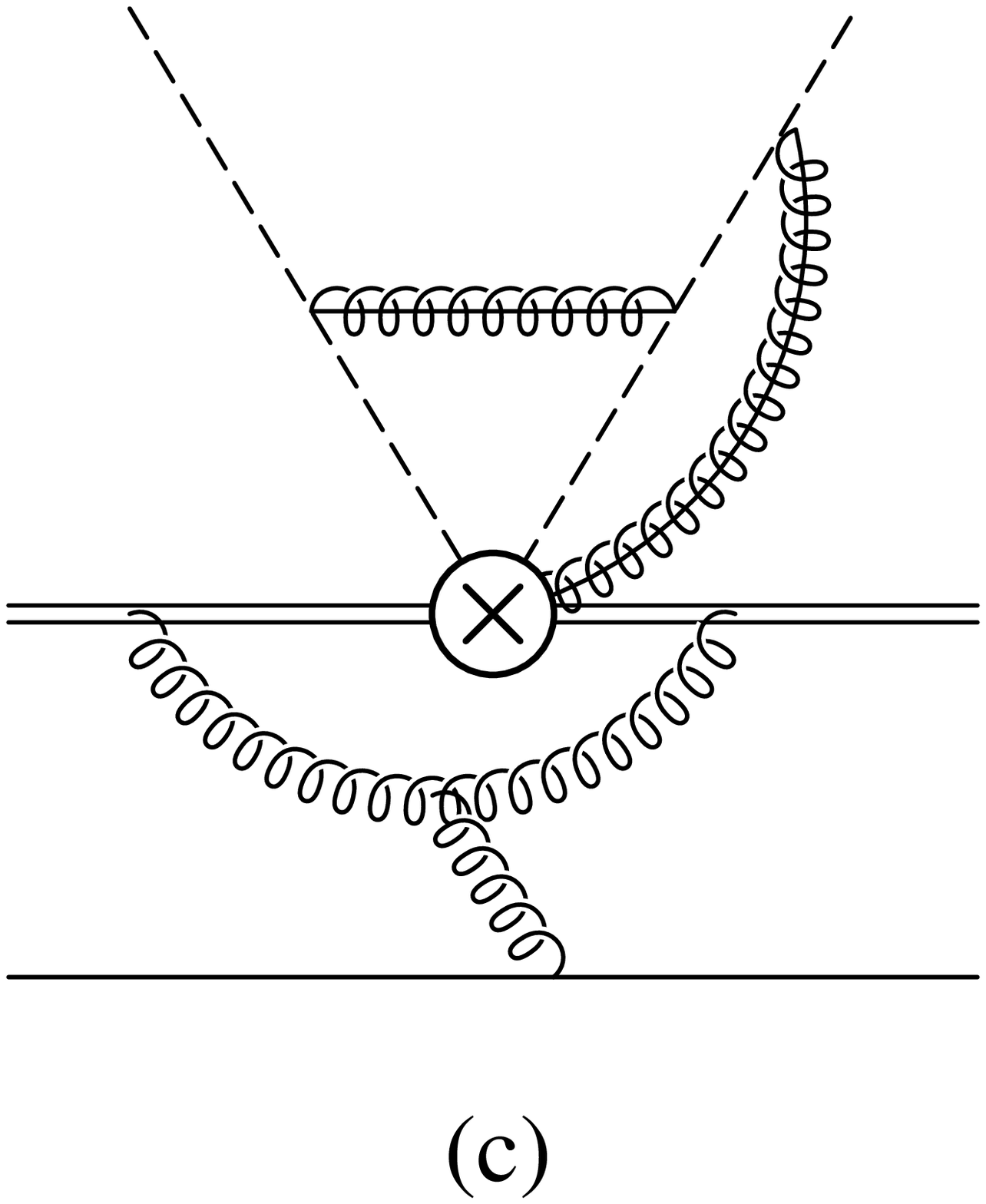}
 }
\begin{picture}(0,0)
 \put(5,78){${\cal L}^{(2)}_c$}
 \put(8,90){\psline[linewidth=2pt]{->}(-0.2,0.6)}
\end{picture}
\vspace{-1.5cm} \caption{Graphs used for the power counting examples in the
text. (a), (b), and (c) correspond to \\[-5pt] the processes DIS,\: $B\to X_u
\ell \bar\nu_\ell$,\: and $B\to D\pi$\: respectively.\ \ Dashed lines are
collinear quarks,\\[-5pt] double solid lines are usoft or soft heavy quarks, and
the single solid lines are soft light quarks.\\[-5pt]  Gluons with a line
through them are collinear, while those without a line are soft or usoft.
\label{fig_1}}
\end{figure}

In the direct method of power counting graphs one counts powers of $\lambda$ for
the loop measures, propagators, and vertices. As our first application we use
the graphs in Fig.~\ref{fig_1} to contrast the direct formula in
Eq.~(\ref{direct}) with the simpler vertex formula in Eq.~(\ref{pc}).  As
mentioned above, in the direct approach the choice of gauge can move powers of
$\lambda$ between the propagators and vertices, and for simplicity we will use
Feynman gauge for our examples. On the other hand, the formula in
Eq.~(\ref{pc}) is applied the same way in any gauge.

The one-loop graph in Fig.~\ref{fig_1}(a) involves an insertion of the DIS
operator in Eq.~(\ref{DIS}) and two leading order interactions from the
collinear Lagrangian ${\cal L}_c^{(0)}$ in Eq.~(\ref{Lc}). Using the vertex
power counting formula in Eq.~(\ref{pc2}), we simply count $V_2^C=1$ and
$V_4^C=2$ to give $\delta=4+(2-4)+0=2$. Alternatively, in the direct power
counting approach we count powers of $\lambda$ for all the loop measures,
propagators, vertices, and external lines. For Fig.~\ref{fig_1}(a) we then have
\begin{eqnarray} \label{dpc1}
 (\lambda)^2 \Big[  \lambda^4 \times \Big(\frac{1}{\lambda^2}\Big)^3 
   \times \lambda^2 \Big] = \lambda^2 \,.
\end{eqnarray}
Here the factor outside the square brackets is for the external fields, the
first term in the square brackets counts the collinear loop measure, and the
second factor counts the three collinear propagators. The last factor in the
bracket counts the powers of momentum in the quark-quark-gluon vertices in
${\cal L}_c^{(0)}$, which in Feynman gauge are either $(\perp, \perp) \sim
(\lambda, \lambda)$ or $(\bn, n) \sim (\lambda^2, \lambda^0)$. The result in
Eq.~(\ref{dpc1}) is again $\delta=2$. Since $\delta=2$, this graph is the same
order as the operator ${\cal O}_1$ itself, and non-perturbative collinear gluon
exchanges such as the one in Fig.~\ref{fig_1}(a) contribute to building up the
parton distribution function.

Next we consider an example involving power suppression for the inclusive decay
$B\to X_u\ell\bar\nu_\ell$ in the endpoint region where $p_X^2 \sim
m_B\Lambda_{\rm QCD}$ and $\lambda=\sqrt{\Lambda_{\rm QCD}/m_b}$. At leading
order the $b$ to $u$ weak current matches onto the SCET currents~\cite{bfps}
\begin{eqnarray} \label{Ji}
  J_i^\mu &=& \big[\, C_i(\bnP,\mu)\: \bar \xi_{n,p}\, W\, \Gamma_\mu^{C_i} 
  \, h_v\, \big]
\end{eqnarray}
where $i=1,2,3$ and $\Gamma_\mu^{C_{1,2,3}}= P_{R} \{\gamma_\perp^\mu\,,
n^\mu\bn\mcdot v\,, v^\mu \}$ with $P_{R,L}=(1\!\pm\!\gamma_5)/2$. In
Eq.~(\ref{Ji}) the heavy quark field $h_v$ is usoft and coefficients $C_i$ are
dimensionless.  Counting factors of $\lambda$ we find that each $J_i^\mu$ counts
as $V_4^C=1$. Using Eq.~(\ref{pc}) we see that $\delta=4$ is the base
$\lambda$-dimension for the tree level time-ordered product of two $J_i^\mu$
currents. With the direct counting method the $\delta=4$ result is obtained by
counting $\lambda^6$ for the external heavy quark lines and $\lambda^{-2}$ for
the collinear quark propagator.  We will also consider currents contributing to
$B\to X_u\ell\bar\nu_\ell$ that are suppressed by a single power of $\lambda$:
\begin{eqnarray} \label{J12}
   {\cal O}_{i\,\mu} &=& \Big[\,B_i(\bnP,\mu)\ \bar\xi_{n,p} \: 
  \frac{\bnslash}{2}\,\big( \ppP^{\dagger\alpha} + g A_{n,q}^{\perp\alpha}\big) 
  W \: \frac{1}{\bnP^\dagger}\:  \Gamma^{B_i}_{\alpha\mu}\: h_v 
    \, \Big]\,, \nn \\[4pt]
   {\cal K}_{i\,\mu} &=& \Big[\, E_i(\bnP,\mu)\ \bar\xi_{n,p} \:
     \Gamma^{E_i}_{\mu\alpha}\,  
    \big( \ppP^\alpha + g A_{n,q}^{\perp\alpha} \big) 
    W \, \frac{1}{m_b}\: \frac{\nslash}{2}\: h_v \,\Big] \,, 
\end{eqnarray}
with dimensionless Wilson coefficients $B_{i}$ and $E_i$, and Dirac structures
$\Gamma_{\alpha\mu}^{B_{1,2,3,4}}=P_L \{\gamma_\alpha \gamma^\perp_\mu\,,
\gamma_\alpha n_\mu \bn\mcdot v\,, \gamma_\alpha v_\mu\,, g_{\mu\alpha}\}$ and
$\Gamma_{\mu\alpha}^{E_{1,2,3,4}}= P_R\, \bnslash
\{\gamma^\perp_\mu\gamma_\alpha \,, \gamma_\alpha n_\mu \bn\mcdot v\,,
\gamma_\alpha v_\mu\,, g_{\mu\alpha}\}$.  Matching the current $\bar u
\gamma^\mu P_L b$ at tree level gives $B_1=1$, $B_2=1$, and $E_2=1/2$.  Each of
${\cal O}_{i\,\mu}$ and ${\cal K}_{i\,\mu}$ count as $V_5^C=1$. The currents
${\cal O}_{i\,\mu}$ were first introduced in Ref.~\cite{chay}, while the
currents ${\cal K}_{i\,\mu}$ are new.\footnote{In Ref.~\cite{chay} it was shown
that reparameterization invariance (RPI) uniquely fixes the $B_i$'s in terms of
$C_i$'s (this was referred to as type-I RPI in Ref.~\cite{mmps}).  It is easy to
show that the ${\cal K}_i^\mu$ currents are not connected to $J_i^\mu$ by type-I
RPI since $\delta_{\rm I}\: \big( \SppP + g \Aslash_{n,q}^\perp\big) W\ \propto\
\bn\mcdot D_c W=0$ so that $\delta_{\rm I} {\cal K}_{i\mu} = {\cal
O}(\lambda^6)$.}

In Fig.~\ref{fig_1}(b) we show a graph contributing to the forward scattering
amplitude for $B\to X_u\ell\bar\nu_\ell$ with the current ${\cal O}_{1\mu}$ on
the left and the current ${\cal K}_{2\mu}$ on the right.  The remaining vertices
in the graph are from the lowest order Lagrangians, except for the one labeled
${\cal L}^{(2)}_c$ which we take from the $\bn\mcdot A_{us}$ in the $(\bn\cdot
iD)$ term in Eq.~(\ref{Lc}).  Now for the loop graph in Fig.~\ref{fig_1}(b) we
have $V_4^C=1$, $V_5^C=2$, $V_8^U=1$, and for the ${\cal L}_c^{(2)}$ vertex
$V_6^C=1$ so Eq.~(\ref{pc}) gives $\delta=4+2(5-4)+(6-4)=8$. Thus, due to the
insertion of a vertex from ${\cal L}_c^{(2)}$ and the subleading currents the
graph is suppressed by $\lambda^4$ relative to leading order diagrams. In
contrast with the direct power counting method we have
\begin{eqnarray}
  \lambda^6\: 
 \ \Big[\lambda^8\times\frac{1}{\lambda^2\,\lambda^4} \times\lambda^2\Big]
 \ \Big[ \lambda^4 \times \Big(\frac{1}{\lambda^2}\Big)^4 \times \lambda^0
  \Big] \big[ \lambda^1 \lambda^1 \big] = \lambda^8 \,.
\end{eqnarray}
Here the first term counts the dimension of the external heavy quark fields.
The factors in the first square bracket are the usoft measure, propagators, and
vertices respectively. In the second square bracket we give the $\lambda$
factors for the collinear loop measure, four collinear propagators, and
collinear vertices (in Feynman gauge). In the final square bracket we have the
powers of $\lambda$ from the currents. The total result $\delta=8$ is the
same as with the vertex formula.

The last diagram in Fig.~\ref{fig_1} involve the weak flavor changing operator
for the non-leptonic decay $B\to D\pi$. At leading power the external operator
in SCET has the form~\cite{cbis,bps}
\begin{eqnarray}\label{Q08fact}
 Q_{\bf \{0,8\}} &=&  \Big( \bar h_{v'}^{(c)}S \Gamma_h \:\{{\bf 1,T^A}\} 
  S^\dagger h_v^{(b)} \Big) \Big( \bar \xi_{n,p'}^{(d)} W 
  C_{\bf \{0,8\}}(\bnP,\bnPd) 
  \, \Gamma_\ell\:\{{\bf 1,T^A}\} \, W^\dagger\,    \xi_{n,p}^{(u)} \Big) \,,
\end{eqnarray}
where $h_{v'}$ and $h_v$ are soft HQET fields and $\Gamma_{h,\ell}$ are spin
structures.  From Table~\ref{table_pc} we see that $Q_{\bf \{0,8\}}\sim
\lambda^5$, and because of the presence of both soft and collinear fields it
counts as $V_5^{SC}=1$. Thus, $\delta=5$ is the base $\lambda$-dimension for
this process.  In the three loop graph in Fig.~\ref{fig_1}(c) all interactions
except $Q_{\bf 0}$ are taken from the lowest order Lagrangians. Here the
advantage of the vertex power counting is more clear. Applying Eq.~(\ref{pc2})
to this graph we see that $V_5^{SC}=1$, $V_4^C=3$, and $V_4^S=4$ so that
$\delta=4+1=5$ and the graph is leading order (since $Q_1\sim\lambda^5$). The
direct counting is more involved giving
\begin{eqnarray} \label{dpc3}
  \lambda^5\: 
 \Big\{ \frac{(\lambda^{3/2})^2}{\lambda^2}\Big\} 
 \ \Big[\lambda^4\times\frac{1}{(\lambda)^2\,(\lambda^2)^2} \times\lambda\Big]
 \ \Big[ (\lambda^4)^2 \times \Big(\frac{1}{\lambda^2}\Big)^5 \times \lambda^2
  \Big] = \lambda^5 \,.
\end{eqnarray}
Here the first term counts the dimension of the external heavy quark fields and
collinear quark fields. The term in curly brackets counts powers of $\lambda$
from the light soft spectator quark lines and the soft gluon propagator that
does not participate in a loop. The factors in the first square bracket are the
measure, propagators, and vertices for the soft loop. In the final square
bracket we give the $\lambda$ factors for the measures, propagators, and
vertices in the two collinear loops (in Feynman gauge). The final $\delta=5$
result is of course the same. Note that the diagram in Fig.~\ref{fig_1}(c) is
also order $\delta=5$ without the spectator interaction. With the vertex power
counting this is obvious since we simply have fewer soft vertices ($V_4^S=2$)
which does not affect the result for $\delta$. With the direct power counting we
must go back and adjust the analysis in Eq.~(\ref{dpc3}) to find this result.

From Eq.~(\ref{pc}) it should be clear that the complete set of leading order
graphs can be constructed by simply adding any number of $V_4^C$, $V_4^S$, and
$V_8^{U}$ interactions. This is in fact the true strength of this power counting
formula, it makes determining the set of all graphs that contribute at a given
order quite simple. 

\vspace{0.2cm}
\noindent{\bf Application 2: Counting powers of $Q$} 
\vspace{0.2cm}

As our second application we show how Eq.~(\ref{pc}) plus dimensional analysis
can be used to determine the power of $Q$ of matrix elements in
SCET. Essentially, a power of $Q$ is assigned by dimensional analysis and then
the power of $\lambda=\Lambda_{\rm QCD}/Q$ determines how many $Q$'s are turned
into factors of $\Lambda_{\rm QCD}$.  Thus, in general a matrix element $\langle
{\cal O}\rangle$ of mass dimension $d$ is order $Q^q$, where $q= d-\delta$
and $\delta$ counts the $\lambda$-dimension of the operator and states.  We use
relativistic normalization for our states, $\langle p' |p\rangle = 2p^0\:
\delta^3({\bf p}\!-\!{\bf p'})$, and find that collinear protons and pions have
$|p_n\rangle \sim |\pi_n\rangle \sim Q^{-1}\lambda^{-1}$ while soft B or D
mesons have $|B_v \rangle\sim |D_{v'}\rangle \sim Q^{-1}\lambda^{-3/2}$.

For DIS, the operator in Eq.~(\ref{DIS}) involves $\xi\sim Q^{3/2}\lambda$ and
$W\sim Q^0\lambda^0$, plus a dimensionless Wilson coefficient, so ${\cal
O}_1\sim Q^2\lambda^2$. Counting powers in the collinear proton matrix element
\begin{eqnarray}
  \langle p_n | {\cal O}_1 | p_n \rangle \sim \big[Q^{-1}\lambda^{-1}\big] \
   \big[ Q^2 \lambda^2 \big] \big[Q^{-1}\lambda^{-1} \big]
    = Q^0 \lambda^0 \,,
\end{eqnarray}
so the result is dimensionless. This agrees with the fact that the proton matrix
element of ${\cal O}_1$ is simply a convolution of a dimensionless hard
coefficient and the quark parton distribution function. 

For $B\to D\pi$ the operator $Q_{\bf 0}$ involves $\xi$, $W$, and $h_v\sim
Q^{3/2}\lambda^{3/2}$ fields, and scales as $Q_{\bf 0}\sim Q^6\lambda^5$.
Taking the $B\to D\pi$ matrix element then gives
\begin{eqnarray}
 \langle D\pi_n | Q_{\bf 1} | B \rangle \sim \big[ \lambda^{-5/2}Q^{-2} \big]\:
  \big[ \lambda^5 Q^6 \big]\big[ \lambda^{-3/2}Q^{-1}\big]
  = Q^3 \lambda  = Q^2 \Lambda_{\rm QCD}  \,.
\end{eqnarray}
This agrees with the fact that this matrix element can be calculated to give a
product of the $B\to D$ form factor and a convolution with the light cone pion
wavefunction. In this result the dimensions are given by $m_B E_\pi f_\pi\sim
Q^2\Lambda_{\rm QCD}$ (see for example Ref.~\cite{bps}).

The SCET formalism can also be used to give a simple derivation of the $Q$
scaling of the hard scattering component of the electromagnetic form factor of
an arbitrary hadron, $\gamma^* h(p)\to h'(p')$.  We assume that the form factor
involves the hard interaction of $k$ collinear partons in both $h$ and $h'$. In
this case the electromagnetic current is matched onto an operator ${Q^{3-3k}}\:
C\: {\cal O}^{(k)}_\mu$, which contains $k$ $\xi_{\bn}$ fields, $k$ $\bar \xi_n$
fields, and $C$, a dimensionless Wilson coefficient. Note that ${Q^{3-3k}}\:
{\cal O}^{(k)}_\mu$ has overall mass dimension $3$. The matrix element of this
SCET current is
\begin{eqnarray}\label{h2h'}
  \frac{1}{Q^{3k-3}} \langle h'_n|  {\cal O}^{(k)}_\mu |h_n\rangle 
  \sim \frac{1}{Q^{3k-3}} [Q^{-1}\lambda^{-1}][Q^{3k}\lambda^{2k}]
   [Q^{-1}\lambda^{-1}]
  = Q \lambda^{2k-2} = \frac{(\Lambda_{\rm QCD})^{2k-2}}{Q^{2k-3}}\,.
\end{eqnarray}
This scaling for the matrix element of the hard scattering component of the
electromagnetic current agrees with the scaling law derived by Brodsky and
Farrar~\cite{BF}.

\vspace{0.2cm}
\noindent{\bf Application 3: Relation to Twist in DIS  }
\vspace{0.2cm}

In the application of the operator product expansion (OPE) to inclusive
processes it is useful to classify contributions according to twist. Two common
definitions are {\em geometric twist}, $\tau=d-s$, equal to dimension minus spin
for an operator, and {\em dynamic twist}, $t$, defined by the $Q$ scaling by
which the matrix element of an operator contributes to the cross
section~\cite{Jaffe}.  At lowest order the definitions coincide,
$t=\tau=2$. However, beyond twist-2 one has $t\ge \tau$, which corresponds to
the fact that operators of a given geometric twist also contribute to higher
orders in $1/Q$ in the cross section due to Wandzura-Wilczek
contributions~\cite{WW}.  Since the $\lambda$-dimension, $\delta$, of operators
also determines their order in $\Lambda_{\rm QCD}/Q$, one might expect that
$\delta$ and twist are related. In fact, dynamic twist $t=\delta$.

As a simple example consider the DIS operator in Eq.~(\ref{DIS}). Recall that
collinear fermion fields have twist $\tau=3/2-1/2=1$, which from
Table~\ref{table_pc} is equal to their $\lambda$-dimension. In the Breit frame
the proton momentum $\bn\cdot p=Q/x$ and the matrix element~\cite{bfprs}
\begin{eqnarray}
  \langle p_n | {\cal O}_1 | p_n\rangle 
 &=& \frac{1}{x} \int_x^1 \!\!d\xi \, C_1\Big(\frac{2Q\xi}{x},0,Q,\mu\Big)\, 
   \big[ f_{i/p}(\xi) + \bar f_{{i}/p}(\xi) \big] \,,
\end{eqnarray}
where $f_{i/p}(\xi)$ and $\bar f_{{i}/p}(\xi)$ are spin independent quark
and antiquark distribution functions. These distribution functions are related
to proton matrix elements of the local twist-2 operators
\begin{eqnarray}
 {\cal S}\ \Big\{ \bar\xi_{n,p'} \gamma^\nu\: 
  i\tensor D_c^{\mu_1} \cdots i\tensor D_c^{\mu_k} \xi_{n,p}\Big\} \,,
\end{eqnarray}
where ${\cal S}$ takes the symmetric traceless combination of the indices $\nu$,
$\mu_1, \ldots, \mu_k$. It is easy to see that the operator ${\cal O}_1$ in
Eq.~(\ref{DIS}) corresponds to the $t=2$ part of this $\tau=2$ tower of
completely local operators,
\begin{eqnarray}
 {\cal O}_1 &=& \sum_{k=0}^{\infty} c_k \, {\cal O}_1^{(k)} \,,
 \qquad\quad 
 {\cal O}_1^{(k)} = \frac{\bn_\nu \bn_{\mu_1}\cdots \bn_{\mu_k}}{2\,Q}\: 
    \bar\xi_{n,p'} \gamma^\nu\: 
  i\tensor D_c^{\mu_1} \cdots i\tensor D_c^{\mu_k} \xi_{n,p} \,,
\end{eqnarray}
where $C_1(z,0,Q,\mu)=\sum_k c_k\, z^k$ and $iD_c^\mu = \cP^\mu+ i\bn^\mu
n\mcdot\partial/2 + g A_{n,q}^\mu $ are collinear covariant derivatives. The
tensor product $\bn_\nu\bn_{\mu_1}\cdots\bn_{\mu_k}$ is symmetric and traceless
and picks out the leading symmetric traceless part out of the local operator.
Thus, we see that in this example that the twist $t=\tau=2$ is identical to the
$\lambda$-dimension of ${\cal O}_1$ which is $\delta=2$.

Next consider the matrix elements of higher order DIS operators.  In Application
2 it was shown that the $\lambda$-dimension of an operator has a direct
correspondence with the power of $Q$ in its matrix element. Thus, it is easy to
see that a mass dimension-2 operator ${\cal O}_i$ with $\lambda$ dimension
$\delta$ has $\langle p_n| {\cal O}_i | p_n \rangle \sim Q^{2-\delta}$, so that
$t=\delta$ even beyond twist-2. Beyond twist-2 the relationship between $\delta$
with $\tau$ is complicated by the same features that complicate the relationship
between $t$ and $\tau$ (for a more detailed discussion see
Refs.~\cite{WW,Jaffe,Geyer}). Thus, there is no simple correspondence between
$\lambda$-dimension and geometric twist. Finally note that the
$\lambda$-dimension also classifies operators in situations where there is no
OPE such as for exclusive decays.

\vspace{0.2cm}
\noindent{\bf Conclusion}
\vspace{0.2cm}

In this paper we have given a detailed derivation of the SCET vertex power
counting formula in Eq.~(\ref{pc}), and showed its equivalence with the more
intuitive but more tedious direct power counting method.  The vertex formula and
$\lambda$-dimension of operators make counting powers of $\Lambda_{\rm QCD}/Q$
in interactions between soft and collinear particles as simple as counting
powers of $1/m$ in HQET. Three examples highlighting the power and simplicity of
the vertex power counting were then given using the processes of DIS, $B\to
X_u\ell\bar\nu_\ell$, and $B\to D\pi$. As a further application we showed how
the vertex power counting can be used to determine the powers of $Q$ and
$\Lambda_{\rm QCD}$ to associate with matrix elements and reproduced the Brodsky
and Farrar result for the $Q$ scaling of the hard scattering electromagnetic
form factor of a hadron. For deep inelastic scattering we then compared the
vertex power counting approach with twist power counting. Finally, we would like
to emphasize that the most important utility of Eq.~(\ref{pc}) is to counting
powers of $\lambda$, $Q$, or $\Lambda_{\rm QCD}$ for power corrections to
processes where a priori even the size of matrix elements may otherwise be
unknown.

\begin{acknowledgments}
This work was supported in part by the Department of Energy under the grants
DOE-FG03-97ER40546 and DE-FG03-00-ER-41132.
\end{acknowledgments}

\newpage

\end{document}